\documentclass[sigconf]{acmart}
\usepackage[utf8]{inputenc}
\usepackage{graphicx}
\usepackage{longtable}
\usepackage{hyperref}
\usepackage{caption}

\usepackage{graphicx}
\graphicspath{ {./supplements/} }

\newcommand{\rqone}{What experiences influence software engineer's self-efficacy?} 

\newcommand{\rqtwo}{How self-efficacy impact software engineers' behavior? }

\newcommand{\rqthree}{How self-efficacy impacts the outcome of work in software engineering teams?}

\newcommand{\participantes}{31 }
\newcommand{\horas}{20 }
\hypersetup{
colorlinks=true,
linkcolor=blue,
filecolor=magenta,      
urlcolor=cyan,
pdftitle={Overleaf Example},
pdfpagemode=FullScreen,
}

\title{Understanding Self-Efficacy in the Context of Software Engineering: A Qualitative Study in the Industry}
\begin{document}
\author{Danilo Monteiro Ribeiro}
\email{danilo.ribeiro@zup.com.br}
\affiliation{%
  \institution{Zup Innovation}
  \city{São Paulo}
  \state{São Paulo}
  \country{Brazil}
}
\affiliation{%
  \institution{Faculdade Senac}
  \city{Recife}
  \state{Pernambuco}
  \country{Brazil}
}

\author{Rayfran Rocha Lima}
\email{rrl2@cesar.school}
\affiliation{%
  \institution{Cesar School}
  \city{Recife}
  \state{PE}
  \country{Brazil}
}

\author{César França}
\email{franssa@cesar.school}
\affiliation{%
  \institution{Cesar School}
  \city{Recife}
  \state{PE}
  \country{Brazil}
}
\affiliation{%
  \institution{UFRPE}
  \city{Recife}
  \state{PE}
  \country{Brazil}
}
\author{Alberto de Souza}
\email{alberto.tavares@zup.com.br}
\affiliation{%
  \institution{Zup Innovation}
  \city{São Paulo}
  \state{São Paulo}
  \country{Brazil}
}
\author{Isadora Cardoso-Pereira}
\email{isadora.silva@zup.com.br}
\affiliation{%
  \institution{Zup Innovation}
  \city{São Paulo}
  \state{São Paulo}
  \country{Brazil}
}
\author{Gustavo Pinto}
\email{gustavo.pinto@zup.com.br}
\affiliation{%
  \institution{UFPA \& Zup Innovation}
  \city{Belém}
  \state{Pará}
  \country{Brazil}
}

\begin{abstract}
\textbf{CONTEXT}: Self-efficacy is a concept researched in various areas of knowledge that impacts various factors such as performance, satisfaction, and motivation. In Software Engineering, it has mainly been studied in the academic context, presenting results similar to other areas of knowledge. However, it is also important to understand its impact in the industrial context. \textbf{OBJECTIVE}: Therefore, this study aims to understand the impact on the software development context with a focus on understanding the behavioral signs of self-efficacy in software engineers and how self-efficacy can impact the work-day of software engineers. \textbf{METHOD}: A qualitative research was conducted using semi-structured questionnaires with 31 interviewees from a software development company located in Brazil. The interviewees participated in a Bootcamp and were later assigned to software development teams. Thematic analysis was used to analyze the data. \textbf{RESULTS}: In the perception of the interviewees, 21 signs were found that are related to people with high and low self-efficacy. These signs were divided into two dimensions: social and cognitive. Also, 18  situations were found that can lead to an increase or decrease of self-efficacy of software engineers. Finally, 12 factors were mentioned that can impact software development teams. \textbf{CONCLUSION}: This work evidences a set of behavioral signs that can help team leaders to better perceive the self-efficacy of their members. It also presents a set of situations that both leaders and individuals can use to improve their self-efficacy in the development context, and finally, factors that can be impacted by self-efficacy in the software development context are also presented. Finally, this work emphasizes the importance of understanding self-efficacy in the industrial context.

\end{abstract}


\copyrightyear{2023}
\acmYear{2023}
\setcopyright{licensedothergov}\acmConference[EASE '23]{Proceedings of the International Conference on Evaluation and Assessment in Software Engineering}{June 14--16, 2023}{Oulu, Finland}
\acmBooktitle{Proceedings of the International Conference on Evaluation and Assessment in Software Engineering (EASE '23), June 14--16, 2023, Oulu, Finland}
\acmPrice{15.00}
\acmDOI{10.1145/3593434.3593467}
\acmISBN{979-8-4007-0044-6/23/06}

\maketitle

\section{Introduction}

Self-Efficacy (SE) is a well-established theory in the social sciences \cite{honicke2016influence,schunk2009self}. Self-Efficacy can be seen as the belief in one's capabilities to organize and execute the courses of action required to produce a given attainment~\cite{bandura1977self}. More concretely, self-efficacy tries to indicate how much an individual~\emph{believes} that s/he is capable of performing a specific task. In the context of software development, when a developer assesses that s/he is able to finish all the tasks by the end of the week, based on their experience in conducting similar tasks, this developer is exercising their self-efficacy. 

Self-efficacy has been the focus of studies in several different areas, such, health, sports, psychology and mainly in education~\cite{friedman2003self, barkley2006reading, mishali2011importance, hamill2003resilience}. According to recent studies, self-efficacy can help to predict one's behavior and also one's performance~\cite{sheeran2016impact}. In the theory, an individual who has \emph{low} self-efficacy tends to avoid difficult tasks or put less effort into completing a task~\cite{artino2012academic}. 

For example, if a developer feels unable to work on a particular section of the code, they might choose not to engage in activities related to that part or may experience insecurity regarding their tasks. In a similar vein, if an individual doubts their ability to learn a new technology within the required timeframe for project implementation, they may put forth less effort in accomplishing that task.

In the context of Software Engineering, we can found several studies about Self-efficacy. Most of these studies investigated the students' perception of their self-efficacy.  
For example, Moores and Chang~\cite{moores2009self} conducted a study with software engineering students and found that overconfidence in self-efficacy is negatively related to performance. Conversely, the authors found that when there was an intervention (e.g., when a mentor gives feedback on student's performance), self-efficacy was positively related to performance. Another study found that pedagogic interventions could improve student's self-efficacy to solve algorithms~\cite{toma2018self}.
Focusing on an industrial setting, França~\cite{franca2009empirical} seek to understand the behaviors of individuals with high/low levels of self-efficacy and what are the triggers of high or low self-efficacy. These studies provide initial evidence that self-efficacy could impact software development practices.

Despite its importance and the promising findings observed, there are few studies that observed the potential impact of self-efficacy in the context of software development. 
Understanding how self-efficacy could be used as a tool in the software engineering arsenal could be useful in many scenarios, for instance, 1) software development team leaders could anticipate situations that may impact self-efficacy and, thus, better manage their teams;  
2) software developers could work to improve their self-efficacy, and thus better understand themselves and have a more successful career.

This study is part of the field of Behavioral Software Engineering \cite{lenberg2015behavioral} and aims to investigate the software developers perception about the potential impact of self-efficacy on their day-to-day activities.
In particular, we are interested in understanding what are the impacts of the different levels (high/low self-efficacy) of self-efficacy in software engineers' behavior, what are the main factors that can affect software engineers' self-efficacy levels, and what are the impacts of self-efficacy levels in software development team.
To shed some light on these questions, we conduct an interview-based study in a large Brazilian software producing organization. We interviewed \participantes employees that participated in a career accelerating program conducted at the company. 

In the perception of the interviewees, 21 signs were found that are related to people with high and low self-efficacy. These signs were divided into two dimensions: social and cognitive. Also, 18  situations were found that can lead to an increase or decrease of self-efficacy of software engineers. Finally, 12 factors were mentioned that can impact software development teams.


\section{Self-efficacy Theory}


Self-efficacy is proposed as a construct of Social Cognitive Theory~\cite{bandura1977social} and can be defined as the belief in one's own ability to accomplish something successfully~\cite{bandura1977self}. 
Self-efficacy can change the perception of reality and how individuals behave~\cite{bandura2010self}, the main assumption of self-efficacy is that people generally will only attempt things they believe they can accomplish and will not attempt things they believe they will fail~\cite{bandura1977self}. This is because individuals generally do not put effort to do a task that they believe they cannot do~\cite{bandura1977self}. When individuals have a strong sense of efficacy, they trust that they can accomplish even the most difficult tasks. Furthermore, they can see the task as challenges to be mastered, rather than threats to be avoided~\cite{bandura1994ramachaudran}.

Bandura affirms that there are four sources of self-efficacy.  The first and most effective way to impact self-efficacy is by mastering experiences~\cite{bandura1994ramachaudran}. The process of mastering an experience is based on doing a task and succeeding in doing it. This occurs because individuals tend to believe that they are more capable to do a task if they have performed well a similar task before.

A famous example of mastery experience is babysitting. Women who have experience taking care of kids before becoming mothers are more confident with their abilities to take care their kids~\cite{froman1989infant}.

Therefore, to master at something, an individual has to practice the task. However, Bandura~\cite{bandura1994ramachaudran} presents some important points. For example, if the tasks are easy or very similar to the ones already performed, self-efficacy may be poorly developed by individuals. Moreover, the task needs to be difficult but not impossible. The individual needs to approach difficulty tasks and work through obstacles to improve self-efficacy~\cite{bandura2010self}. 
Heslin~\cite{heslin1999boosting} gives another tip to improve mastery experience: breaking down a difficult task into small steps, which are relatively easy, to ensure a high level of initial success, then progressively increase the difficulty of the task so that the participant still feels challenged. At last, Heslin suggests providing feedback through workshops, training programs, internships, and clinical experiences to improve mastery experiences. 

Another factor is vicarious experience. Vicarious experience can occur when an individual observed others' successes and failures who are similar to him/her. Bandura~\cite{bandura1977self} mentioned that watching someone like you (e.g., a colleague in a similar level of experience) doing and successfully accomplishing something you would like to do (e.g., a task during the sprint) can increase \emph{your} self-efficacy. Conversely, when you are observing someone like you to fail, \emph{your} self-efficacy tends to be negatively impacted. However, failure is not always harmful; when they feel confident, individuals can avoid repeating the errors they observed others doing~\cite{brown2013self}. 
Heslin~\cite{heslin1999boosting} also commented that to improve self-efficacy, a manager can do workshops and training sessions because when individuals are watching others in a training session, the manager can provide observational experiences.
 
The third factor affecting self-efficacy is verbal persuasion. When individuals are persuaded verbally, they may become more confident that they can do the task~\cite{bandura1977self}. Having others verbally supporting the attainment of a task helps to support a person's belief in himself or herself~\cite{bandura1977self}. 
Verbal persuasion is frequently used by sports coaches to improve self-efficacy~\cite{brown2013self}. Verbal persuasion builds self-efficacy when managers encourage and praise their competence and ability to improve their performance~\cite{heslin1999boosting}. 
 
The fourth and last factor is physical and emotional states that occur when someone contemplates doing something~\cite{bandura1977self}. For example, anxiety, fear, perception and worry can affect self-efficacy negatively and can lead an individual to believe that she cannot perform a task~\cite{pajares1996self}.

\subsection{Self-efficacy in Software Engineering}

Some studies investigated self-efficacy  in Software Engineering. Tsai and Cheng~\cite{tsai2010programmer} conducted an industrial research that found self-efficacy is related positively to intention to knowledge sharing and knowledge sharing. The authors hypothesized that self-efficacy could improve knowledge sharing at software teams.
Another study investigated that the students' programming self-efficacy beliefs had a strong positive impact on the effort and persistence of Software Engineering students. Furthermore, it was found that self-efficacy is negatively related to seeking help~\cite{kanaparan2017self}.

Arya et al.\cite{arya2012moderating} found that self-efficacy is positively but not very significantly related to organizational commitment. Fu~\cite{fu2010information} found that professional self-efficacy is positively related to the career commitment of Software Engineers. 

Hazzan and Seger~\cite{hazzan2010recruiting} identified in their study that high self-efficacy practitioners tend to be: "\textit{more cooperative, have a greater sense of morale working with their team members, feel that their relationships with co-workers are closer, get better managerial support, report higher needs in achievement, dominance, affiliation, and difference and have better attitudes towards change}". 

Other studies are aiming to understand how to improve self-efficacy. For example, Dunlap~\cite{dunlap2005problem} observed that the use of Problem-Based Learning could improve the self-efficacy of Software Engineering students. Steinmacher et al.~\cite{steinmacher2015increasing} found that an online coach called FLOSScoach had a positive influence on open-source newcomers' self-efficacy, making newcomers more confident and comfortable during the project contribution process.
Srisupawong et al.~\cite{srisupawong2018relationship} revealed that perceptions of autonomy, meaningfulness, and involvement are positively associate with strong self-efficacy. Furthermore, the students' perceptions of vicarious experiences and perceptions of social persuasions demonstrated a positive relationship with self-efficacy. Perceived physiological and affective states demonstrated a negative influence self-efficacy of computer science students.

Ribeiro~\cite{ribeiro2022} conducted a survey and found a high negative correlation between self-efficacy about work and Exhaustion. The author also found was negative and weak with technological and team instabilities, and does not exist with task instability. In another study, Ribeiro~\cite{ribeiro2020relaccoes} found that was a high positive correlation between self-efficacy about work and job satisfaction. This is an important result of self-efficacy because it presents self-efficacy as a factor that should be taken into consideration in the context, impacting variables that are more observed both by industry and academia.

\section{Method}

This research focuses on understanding the software developer's perception of self-efficacy's, and its possible impacts on their work. We also want to understand, according to software developers' opinions the follow research questions:

\begin{itemize}
    \item[\textbf{RQ1:}] \rqone 
    \item[\textbf{RQ2:}] \rqtwo 
    \item[\textbf{RQ3:}] \rqthree 
\end{itemize}


First, we want to understand what are the software engineers' \emph{experiences} that could impact self-efficacy (positively or negatively). With RQ1, we intend to generate a list of experiences that can be used by leaders to generate/avoid employee self-efficacy. Second, we want to understand how \emph{self-efficacy} can impact the software engineer's behaviors (RQ2); this is important to clarify what a person with high and low efficacy can do in their day-to-day work. Finally, we want to understand how those behaviors can impact the \emph{work results} (RQ3).

\subsection{Study Design}
We conducted a qualitative interview study~\cite{merriam2015qualitative} to answer these research questions. 
To get their perception, we conducted \participantes semi-structured interviews, containing almost \horas  hours of recorded data. Among the participants, there were 25 men and 6 women. 

This study was carried out within a large Brazilian software development company, consisting of over 3,000 employees. As a consultancy firm, its main emphasis lies in creating software solutions for banking institutions, yet it also engages in software development for other sectors such as telecommunications and services.

It is worth noting that 3 out of 5 authors of this research work at this organization, and this research emerged from organization needs to understand and improve software engineers' performance.


All interviewees are employees (software engineers) and participated in an industrial bootcamp, organized by the company to improve performance of them.
After completing the bootcamp, the participants were assigned to actual software development teams. They were all invited to participate in this study after working on real projects for approximately two months  after bootcamp. Therefore, we employed a convenience sampling approach in selecting potential participants, and each individual had the autonomy to decide whether to participate in the study or not.


\subsubsection{The bootcamp}

The bootcamp can be seen as a career acceleration program and seeks to improve one's path in technology, through an intense and deliberate training perspective. In general, the bootcamps take 12 weeks.

The goal of the bootcamp is to prepare the participants (and company's employees) to handle situations that may occur in the teams to which they will be allocated, arming them with knowledge and tools to solve common real-world problems. Thus, the bootcamp training is based on real situations that are needed to perform daily coding activities in the company. 

At the company, the person can choose whether to work in site or remotely. All participants interviewed work remotely. Although the needed skills vary from bootcamp to bootcamp, overall anyone with some programming skills (e.g., Java, DBMS, and Git) could apply to participate in the bootcamp. The selected participants were hired as software engineers by the company, as we require them to work 40h peer week. 

\subsubsection{The bootcamp team}

A team of experts in the technologies used in the company was recruited to create the bootcamp protocol. These technical mentors  were also responsible for teaching, mentoring, and evaluating the participants. 
We have two technical mentors for each bootcamp. Mentors are also well-known figures in the tech radar in Brazil, including renowned speakers in tech events. 
The mentors are also full-time and dedicated to bootcamp execution. 
Besides the technical mentors, the bootcamp participants are also closely followed by a program manager. The program manager monitors the development of training participants, ensuring the best experience during the program, and supporting the mentor team.

\subsubsection{The bootcamp sessions}

The bootcamp is conducted remotely, with synchronous and asynchronous sessions. The training part occurs mostly in synchronous sessions. The mentors give one hour of a synchronous theoretical class per day. There are also two other synchronous sessions of two hours each during the week. At these sessions, the mentors provided feedback and answer questions. The mentors are available to clarify any doubts during working hours, in a virtual room. Coding sessions such as dojo and pair programming that take place during the training session. 

There are also external materials, such as short courses created by the mentors and other external organizations that students must complete asynchronously to continue advancing in the bootcamp. Finally, for each course, the mentors created exercises for the students to practice.

\subsection{Data Collection}

The data used in this study were collected through semi-structured interview. The interview was conducted in Portuguese, and the script was refined in a pilot study, with three interviewees. These interviewees are part of our close network.
Through this pilot study, we were able to identify questions that were ambiguous, repetitive, or difficult to understand. For example, during the pilot study, we identified the need to explain the concept of self-efficacy to the interviewees in a didactic manner because they were not familiar with the concept. With this in mind, in the subsequent interviews, we had to explain at the beginning of the interviews what is self-efficacy is to the interviewees using the following text: "Self-efficacy is a concept that looks at people's sense of capability in relation to their abilities. People with high self-efficacy feel more capable in relation to their abilities, people with low self-efficacy feel less capable of performing a task when they observe their abilities and compare them with the abilities that the task requires". 

During the interviews, occasionally, during some moments of the interviews, we reinforced this definition, for example: "When you think of someone who has high self-efficacy in relation to their software development abilities, i.e. they feel fully capable of performing a specific task that has been given..." Or: "How do you believe having low self-efficacy, i.e. not feeling capable of performing a task given its requirements, can impact a software developer's career?". We did this in order to avoid any misunderstandings about the term "self-efficacy" which is not commonly encountered in the day-to-day lives of software developers.

The interview script had 12 questions. We asked about the participants' professional trajectory and their experience as a developer as well as their perception about self-efficacy. For example: "How do you think feeling with low self-efficacy about your ability to develop software impacts your team's results?" or "In bootcamp, was there any situation that caused you to feel with low self-efficacy about your abilities to develop software? Tell me more." The full interview script is available in \texttt{https://bityli.com/cxjkP}.

The interviews were conducted in March 2021, and due to COVID restrictions, all interviews were conducted remotely and online using Google Meeting. With the participants' agreement, all videos were recorded in MP4 files and later transcribed by a third party to TXT files before being sent for analysis.

\subsection{Data Analysis}

To perform the data analysis, we chose to use thematic analysis because we want to identify, analyze, and report patterns within the data~\cite{cruzes2011ThematicAnalyse}. Thematic analysis focus in describing and organizing the data in detail. It also helps interpreting different aspects related to the research topic~\cite{cruzes2011ThematicAnalyse}. First, we did an initial reading of the data (TXT files of audio transcription), then we identified specific segments of text and labeled them. We also reduced the overlap between codes by grouping them into themes. Finally, we also combined themes to create higher order themes. This process was supported by ATLAS.ti which is a qualitative data analysis platform, which allows businesses to analyze content including text, graphics, audio, and video for quality.

All analysis was performed in Portuguese. To conduct, the first two authors carried out the process of identification and labeling of the segments of the interviews individually and separately. Afterwards, they compared the results and discussed how the codes, themes and sub-thems would be created.

Whenever conflicts arose, first we tried to resolve it through a research chat group in a assyncron way. If we could not solve it, we held an online resolution meeting with the participation of at least three authors who discussed until achieve a consensus.
During these meetings, we discussed whether to merge or split themes/codes, finding appropriate names for themes/codes, or identifying broad themes at the right level of granularity.

\subsection{Ethics}

As for ethics, during our interviews, we informed the participants that we would not identify them and all data will be  anonymized in this study. In addition, we kept data confidential, including from the participant's company.  Finally, all participants were also informed that although the research originated from the company, they would not face any punishment if they did not want to participate in the research, that they could withdraw at any time, that the results would not be used and observed individually, that they were in a free environment to express their opinion.

We followed the norms of Resolution 466/12 – CNS-MS of the Brazilian National Health Council that regulates research with human subjects. 
These standards define principles that researchers must follow to avoid harming the health of respondents and increase the benefits of research results for participants.

It is worth noting that none of the researchers have worked or are currently working directly with the interviewees.

\section{Results}
\label{sec:resultados}

In this section, we discuss the opinions acquired through interviews with software engineering practitioners that support to answer the research questions.
\subsection{RQ1. \rqone}

\begin{figure}[!ht]
\centering
\includegraphics[scale=0.20]{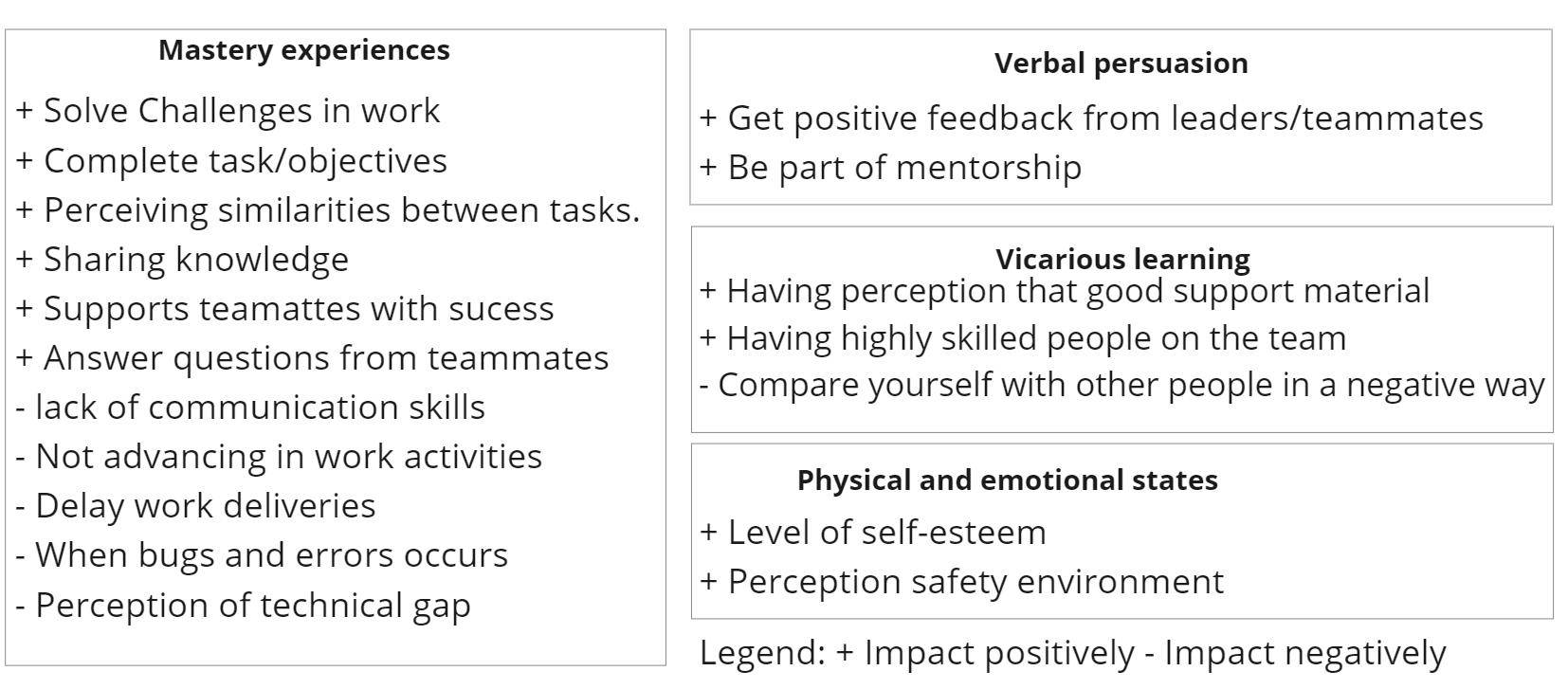}
\vspace{-5mm}
\caption{List of experiences that can impact self-efficacy levels}
\label{fig:rq1}
\vspace{-4mm}
\end{figure}

To answer RQ1, at the end of the thematic analysis we used the division of self-efficacy sources proposed by Bandura \cite{bandura1977self} to group the themes that emerged in the process, namely: Mastery Experiences, Vicarious Experience, Verbal Persuasion, Physical and Emotional States. The final result is presented in see Figure \ref{fig:rq1}.

\subsubsection{Mastery experiences}
This section group all factors that refer to the experience of successfully completing tasks or achieving goals.

\textit{\textbf{Solve Challenges in work}} and \textit{\textbf{complete task/objectives}} were pointed out as a source of self-efficacy.


\begin{quote} \textit{"A situation that made me feel more self-efficacious? I think I can complete the tasks for sure. P002"}
\end{quote}

\begin{quote} \textit{"I was presented with more complicated challenges and I managed to solve them. What increased my self-efficacy was getting results in some difficult tasks. P012"}
\end{quote}

Another situation that can impact self-efficacy is the \textit{\textbf{perception that one is performing an activity that is similar to what they have done before, with success or failure.}}.
 The more you experience similar situations, the more you are and feel able to resolve them. 
 

\begin{quote} \textit{"The bootcamps projects had a lot of similar things between them. I think there was a repetitive factor to them that is important, making you practice more of the same thing...this increased my self-efficacy. P013"}
\end{quote}




Situations like \textit{\textbf{Not advancing in work activities}}, \textit{\textbf{Delay work deliveries}}, and \textit{\textbf{When found bugs and errors}} can decrease self-efficacy.

\begin{quote} \textit{"The fact that I am not making progress is making me feel less self-efficacious. P002"}
\end{quote}

\begin{quote} \textit{"It was my delay in delivering so, I felt that my self-efficacy was decreasing. P003"}
\end{quote}

\begin{quote} \textit{"...Knowing that my solution had problems lowered my self-efficacy...I realized that I need to study more about this. P019"}
\end{quote}


\textit{\textbf{Sharing knowledge and support team}} can increase self-efficacy. When people perceive that they are successfully completing tasks, this can increase their self-efficacy. 

\begin{quote} \textit{"I have some stories where I had just learned a subject[...] five minutes later I got into a video call with a colleague to teach it to him, so I think that increased my self-efficacy. P006"}
\end{quote}

\textbf{\textit{Perception of technical gap}} can decrease self-efficacy.
\begin{quote} \textit{"When I saw the video about Unit Testing, which is something I had never worked on in my life, I thought: Oh my God! What is that? At that moment, I felt my confidence dwindle. P001"}
\end{quote}

\textit{\textbf{Answer questions from teammates}} can increase self-efficacy.

\begin{quote} \textit{"There was a day when someone went to ask a question and then I commented, I gave an example... [about what made him feel more self-efficacious] P021"}
\end{quote}

Our results indicate that perception of \textit{\textbf{lack of communication skills}} can affect self-efficacy and negatively.

\begin{quote} \textit{"The lack of ability to communicate your results or your point of view generates the fear of being misunderstood, which ends up reducing your self-efficacy. P010"}
\end{quote}

\subsubsection{Physical and emotional states}

This section group all factors that refer to the influence of one's physical and emotional states on self-efficacy.


\textit{\textbf{Level of self-esteem}} can influence self-efficacy. We noticed that while people with low self-esteem had more difficulty raising their self-efficacy level, people with high self-esteem, even going through a moment of low self-efficacy, this was a temporary situation and with great chances of being overcome.

\begin{quote} \textit{"Self-efficacy is important not only for a developer career, but for any career in fact. If you don't feel with high self-efficacy, you don't have self-esteem in what you set out to do, you end up giving up on that area. P028"}
\end{quote}

\begin{quote} \textit{"I didn't feel with high self-efficacy, not because I didn't have technical knowledge about something, but because I wasn't psychologically well enough to solve and do the things I needed in life. P009"}
\end{quote}




Promoting an environment of \textit{\textbf{respecting, appreciating, and valuing other team members' contributions}} can increase their self-efficacy.

\begin{quote} \textit{"Being in an environment where apprentices and mentors can exchange ideas about what is better or worse in a given solution gave me more confidence in expressing my opinion. [...] whenever someone commented at the meeting, the leader always tried to value what was said.
This kind of behavior increases not only my self-confidence but other meeting attendants as well. P027"}
\end{quote}


\subsubsection{Vicarious learning}
This section group all factors that refer to observing others successfully completing similar tasks or achieving similar goals.

Our results show that \textit{\textbf{having perception that support material}} can increase self-efficacy.

\begin{quote} \textit{"...there was a very good support material and a well-defined study guide...so, this makes me more "self-efficacy" now. P019"}
\end{quote}

\textit{\textbf{Having highly skilled people on the team}} can increase self-efficacy.

\begin{quote} \textit{"I already knew Jon [leader] through lectures, when I found out he would be on the team it was one of the things that increased my self-efficacy... I think it was one thing... the people who were participating brought me this security, I know these guys here are really good, and I'll be well-supported if there's something I can't solve on my own. P009"}
\end{quote}

\textit{\textbf{Compare yourself with other people in a negative way}} can decrease self-efficacy.

\begin{quote} \textit{"I saw other people moving forward, discussing more about other technologies and I was not able to keep up with them. I realized that I wasn't doing the way I wanted and I was well below expectations. This shook my self-confidence. P017"}
\end{quote}

\subsubsection{Verbal persuasion}

Next, some situations will be presented where individuals received feedback or encouragement from other team members.

\textit{\textbf{Get positive feedback from leaders/teammates}} can increase self-efficacy.

\begin{quote} \textit{"Something that I found cool in the beginning was the feedback provided by the mentors. They showed where we did it right and where we could improve. Having someone with 15 years of experience saying that my code is good, but could be improved, made me more confident in the sense that I feel like I'm improving. P003"}
\end{quote}

\textit{\textbf{Be part of mentorship}} can increase self-efficacy.

\begin{quote} \textit{"Mentoring showed what I'm doing correctly and where I have to improve. Having someone I can talk to and validate my solutions is very important for my confidence. I know I'm not alone. P004"}
\end{quote}


\subsection{RQ2: \rqtwo}

\begin{figure}[!ht]
\centering
\includegraphics[scale=0.36]{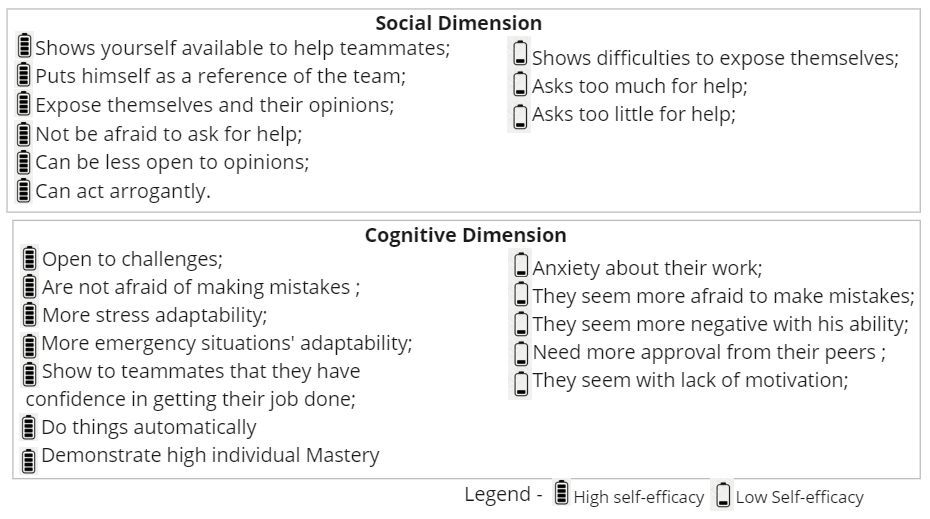}
\vspace{-5mm}
\caption{List of software engineers’ behaviors according to their self-efficacy levels}
\label{fig:rq2}
\vspace{-3mm}
\end{figure}

In this context, the findings are grouped into two dimensions: Social (D1) and Cognitive (D2) as shown in Figure \ref{fig:rq2}. 

\textbf{D1. Social dimension} groups all individual factors found that revolve around how people relate with one another in a team.
The interviewed perceived the individuals with high self-efficacy in software engineering are more \textit{\textbf{available to help teammates}}. 
\begin{quote} \textit{"When I visualize this person with high self-efficacy... I see a willingness to help... P002"}
\end{quote}


Another behavior perceived of individuals with high self-efficacy is that \textit{\textbf{puts himself as a reference of the team}}.

\begin{quote} \textit{"...putting yourself as a reference on a certain thing, an authority on the subject, you end up literally becoming a point of reference for other people. P007"}
\end{quote}


 The interviewees also said that individuals with high self-efficacy are looking to \textit{\textbf{expose themselves and their opinions, and not be afraid to ask for help.}}

\begin{quote} \textit{"...A person [with high self-efficacy] is more likely to expose their opinions about the code... A person with low self-efficacy is somewhat reluctant and thinks the code is wrong. The person with high self-efficacy just talks, even if it's wrong... P012"}
\end{quote}

\begin{quote} \textit{"I notice that a person with high self-efficacy is not afraid to ask for help, that is a differential. P031"}
\end{quote}

Another point of attention is that sometimes individuals with high self-efficacy can be \textit{\textbf{less open to opinions from teammates and can act arrogantly}}.

\begin{quote} \textit{"He [ex-teammate with high self-efficacy] seemed to understand a lot of the things...sometimes he seemed to understand too much, but I think there was a negative point that it seemed that other opinions didn't matter much, I had a feeling, I don't know if that's really what was going through that person's head, but it seemed that he was saying: "I'm above average here". P025"}
\end{quote}


On the other hand, the participants also said that individuals with low self-efficacy are perceived by \textit{\textbf{teammates with problems to expose themselves}}, sometimes, they know the correct answers but don't say to it or have afraid to expose. 
Teammates also notice that they \textit{\textbf{ask too much for help or ask too little for help}}. The first case, can be related with be afraid with make mistakes, be negative and the need to be approval. At second case (ask little help), the individual generally afraid of interfering with other people's work or asking a very simple question.


\begin{quote} \textit{"So, when you have low self-efficacy, you end up taking less initiatives, exposing yourself less...sometime they cannot even look for someone else to ask for help. P001"}
\end{quote}


\begin{quote} \textit{"I think that a person with low self-efficacy is always asking a lot of help from other people. P025"}
\end{quote}


\textbf{D2. Cognitive dimension} group all mental factors perceived by interviewed people. They said individuals with high self-efficacy in software engineering are more \textit{\textbf{open to challenges, are not afraid of making mistakes and show to teammates that they have confidence in getting their job done}}.

\begin{quote} \textit{"Having high self-efficacy motivates me to face new challenges and take responsibility. I am not afraid to expose myself.
    P021"}
\end{quote}

Individuals who appear to have higher self-efficacy also seem to have greater \textit{\textbf{adaptability to stress and emergency situations}}, they better manage work-related stress better and handle uncertain and unpredictable work situations.

\begin{quote} \textit{"The person will feel secure about what they are doing, regardless of being pressured, even with a very tight deadline... P009"}
\end{quote}

\begin{quote} \textit{" [When my self-efficacy is high]...if a problem arises, I do not despair... I know how to solve it easily, and calmly, it's not the kind of thing that I'm going to get desperate... P009"}
\end{quote}


Interviewed also said that high self-efficacy individuals appear to \textit{\textbf{demonstrate high individual Mastery}}.

\begin{quote} \textit{"
... I imagine that the person with high self-efficacy at least has mastery over everything he practices on a day-to-day basis and intends to do. P009"}
\end{quote}
Additionally, some characteristics were cited as negative points which can be found in individuals with high self-efficacy and deserve attention not to harm the team. For example, respondents said that people with high self-efficacy \textit{\textbf{can do things automatically}}, which can increase productivity but can also make people make mistakes.


\begin{quote} \textit{"For example, when you have a lot of self-efficacy in what you do, you may start to relax at certain points. This can be a problem because you are not in attention. P012"}
\end{quote}

They also said that individuals with low self-efficacy look like with more \textit{\textbf{anxiety about their work, are more afraid to make mistakes, are more negative with his ability}}. 

\begin{quote} \textit{"I think people with low self-efficacy are desperate... extremely anxious. I think it's normal to be anxious, but extremely anxious, no. P002"}
\end{quote}

\begin{quote} \textit{"I think it's that person who will always put an obstacle in what she's going to do... she will already say that she won't do it, because it won't work...I think this is a signal of a person with low self-efficacy. P027"
 }
\end{quote}

Individuals with low self-efficacy also \textit{\textbf{need more approval}} from their peers about their work activities, and they \textit{\textbf{are seen with lack of motivation}}.

\begin{quote} \textit{"I think the person with low self-efficacy is always thinking they're wrong, even though they're right. If I do a service, something, it works, I keep thinking: “but what if there's a better way to do it?”; Something like that. Never trust what they are done. P022"}
\end{quote}

\subsection{RQ3. \rqthree}

\begin{figure}[!ht]
\hspace*{-.25cm}
\centering
\includegraphics[scale=0.36]{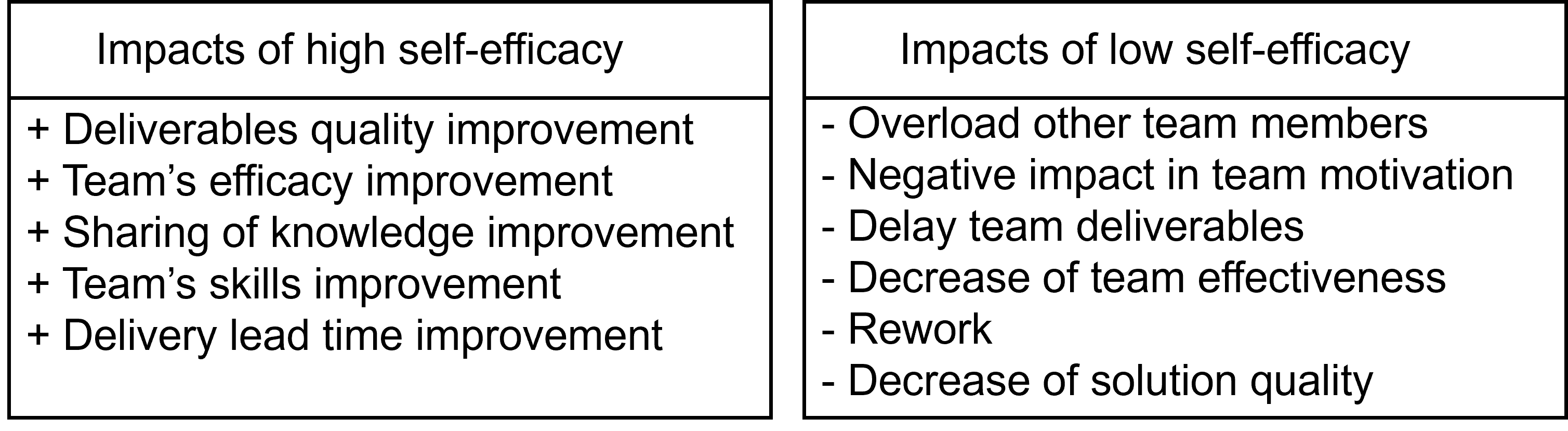}
\vspace{-3mm}
\caption{List of impacts of high and low self-efficacy}
\label{fig:rq3}
\vspace{-3mm}
\end{figure}

This section will present the impacts of behaviors generated by self-efficacy in the software development process. They will be divided into impacts of high self-efficacy and low self-efficacy as shown in Figure \ref{fig:rq3}.

\subsubsection{High self-efficacy}

In the perception of the interviewees, software development teams are impacted by people who have high self-efficacy in different ways. For example, when exposed, people who have high self-efficacy, express their opinion more and take more risks.
This can lead to a \textbf{better quality of the deliverable} because the idea is better discussed in the team.


\begin{quote} \textit{""If I feel more confident in knowing a technology, I will be more confident to suggest, to come and say: "look, maybe, if we do it this way...". With this, the proposed solution may be better... P008"}
\end{quote}

\begin{quote} \textit{" A person with high self-efficacy... I think [...]they produce software with higher quality. P023"}
\end{quote}
Another important point is that people have more faith in people with high self-efficacy, in the opinion of the interviewees, this can make the perception of the \textbf{team's efficacy increase}. Team efficacy is a collective efficacy that occurs when individuals change their perspective from themselves efficacy to the team level \cite{bandura1977self}.

\begin{quote} \textit{"I think this creates trust because you know that if you throw the ball in that person's hands, they will want to solve the problem in the best way possible. P014"}
\end{quote}



According to the interviewees, having individuals with high self-efficacy in the team also helps to have a \textbf{greater sharing of knowledge} within the team, as well as helps to \textbf{increase the overall skills of the team}.

\begin{quote} \textit{"Feeling confident, always helping others, sharing something you know, helps in the growth of others as well, people on your level, people above also, who sometimes don't know something, and especially people below, junior and interns, for example." P015}
\end{quote}



In the perception of the interviewees, individuals with high self-efficacy tend to \textbf{improve the delivery lead time}, e.g., deliver their tasks more quickly. It's worth noting that the fact that they deliver more quickly does not mean that they deliver the best code.

\begin{quote} \textit{"The person who is confident in their code will be able to deliver faster. I'm not saying they won't be careful. They see that the result is correct and they already send it, without being so meticulous in the details of the code. P005"}
\end{quote}

\begin{quote} \textit{"I think they will have better results both in terms of delivering faster and in terms of not being afraid of not being able to do some code. P012"}
\end{quote}




\subsubsection{Low self-efficacy}


According to the interviewees, people with low self-efficacy may \textbf{overload other team members} by not accepting the assignment of some task or transferring unfinished tasks to other team members.

\begin{quote} \textit {"There were tasks that our supervisor would assign to her and she passed it on to someone else because she felt she wasn't capable of doing it... and didn't do it. P026"} 
\end{quote}


Interviewees believe that the fact that an individual has low self-efficacy, can lead them to have negative thoughts about their task and impact the results expected by them, having a \textbf{negative impact on their motivation}.


\begin{quote} \textit{"When it comes to delivering results, the person is not able to deliver the expected results, cannot meet the demand, the team will feel insecure, sometimes the team can also enter the same cycle and get unmotivated." P009}
\end{quote}

 
In the perception of the interviewees, individuals with low self-efficacy can also become insecure and have creative blocks, \textbf{which can delay team deliverables
}.
\begin{quote} \textit{"...it takes longer for her to create blocks in her head and keep pushing, pushing, pushing that thing to do later and that, then it delays, delays everything and gets more nervous, I think that especially in these two reasons."P005}
\end{quote}




Interviewees also believe that \textbf{team efficacy can be negatively impacted} by low self-efficacy of their members.

\begin{quote} \textit{"...sometimes, depending on the number of people [with low self-efficacy], it doesn't impact as much, as for example, if it's only one person among ten... but if there are more people who feel with low self-efficacy, this can impact how much a team feels capable"P014}
\end{quote}

The interviewees point out that due to the fact that individuals do not express their opinions, even if they believe they are correct, it can \textbf{lead to rework} and can impact negatively the \textbf{quality of the solution}.

\begin{quote} \textit{"When the team asked for his opinion, he [a person with low self-efficacy] didn't say anything. Then the meeting ended, and he told me: "the guy chose wrong for such a reason." He didn't have confidence in his ability to question the other on a technical point, but he knew, and what he said was right." P011}
\end{quote}





\section{Discussion}


In this study, we sought a deeper understanding of the role of self-efficacy in the context of the software development industry. We documented \textbf{18 experiences} that modify self-efficacy in the context of Software Engineering, which were classified according to the self-efficacy sources model proposed by Bandura into: Mastery Experiences, Vicarious Experience, Verbal Persuasion, Physical and Emotional States.



According to Bandura \cite{bandura2010self}, the main source of self-efficacy comes from \textbf{Mastery Experiences}. In this study, Mastery Experiences had the highest number (11) of experiences that lead to changing self-efficacy, they are: solve Challenges in work, complete task/objectives, perceiving similarities between tasks, sharing knowledge, supports teammates with success, answer questions from teammates, lack of communication skills, not advancing in work activities, delay work deliveries, When bugs and errors occurs, perception of technical gap.


People's perception of the task is important for building individual's self-efficacy. This fact has also been evidenced in other contexts\cite{heslin1999boosting, mcauley2000self}, but to the best of our knowledge, it is the first time it has been observed in the context of the software development industry.



That being said, managers should monitor the choice of activities of their subordinates, being able to provide challenging tasks aiming to increase their self-efficacy, or even providing similar activities so that the individual can perform and feel more capable. It is also necessary for managers/leaders to be attentive to individuals who do not progress in activities, who are falling behind, whose codes are getting lots of errors and bugs, as these individuals may be having their self-efficacy negatively impacted.


  
In some areas of knowledge such as medicine\cite{nishisaki2007does}, self-efficacy is increased through training and simulations aiming to impact the performance of individuals. In the same direction, Software development companies could also develop training and/or activities to promote individuals' self-efficacy . The use of bootcamps \cite{root2007key}, for example, with activities related to what developers will carry out on their day-to-day, with support from mentors and good materials, may impact the self-efficacy of developers and consequently performance in a general manner.



Good documentation is very important for software development \cite{wilson2017good, haneef1998software}. Our results suggest that having a positive perception of supportive materials, such as books, documentation, and website answers, can also positively influence the self-efficacy. This provides an additional reason for software development companies to produce good documentation and take greater care with the training and materials they provide for their employees.


In some sports like hockey, the skill of group members can impact the self-efficacy of individuals\cite{feltz2001self}. Our results suggest that in Software Engineering, members who perceive a high level of skills among other team members tend to feel more self-efficacious because they can rely on the help of these people if they can't perform the task. That is the main source of \textbf{vicarious experiences}.






Regarding \textbf{verbal persuation}, our results also pointed out to other two possible experiences that can increase an individual's self-efficacy: receiving feedback from leaders/teammates and being part of mentoring. This is consistent with findings from other areas of knowledge \cite{nishisaki2007does,feltz2001self}, and we believe that these practices are already widespread in the software development industry. The key point here is the influence of leadership, not only providing feedback, but also promoting an environment where team members feel encouraged to safely talk to each other.


The last source of self-efficacy is \textbf{physical and emotional states}. Two experiences were found that can activate this source of self-efficacy: levels of self-esteem and perception of safe environment.



In addition, this study aimed to understand how different levels of self-efficacy can influence the behavior of software engineers. We identified two dimensions of behaviors: (i) the \textbf{social} dimension revealed nine impacts (six related to high self-efficacy and three related to low self-efficacy) and (ii) the \textbf{cognitive} dimension revealed twelve impacts (seven related to high self-efficacy and five related to low self-efficacy). However, it is important to note that these behaviors should not be evaluated in isolation since self-efficacy is not the only factor that leads people to develop these behaviors. The existence of these behaviors may serve as a warning sign for leaders and managers to be mindful of the self-efficacy of individuals on their team.


In the \textbf{social dimension}, higher levels of self-efficacy lead individuals to proactively help their co-workers, and to make an individual a reference for the team. These behaviors have a positive impact on the overall performance of the team \cite{salas2017situation, cheah2019mutual}. Expressing opinions and not being afraid to ask questions can be fundamental behaviors in innovative environments \cite{edmondson2006explaining} such as software engineering context.



Nevertheless, two negative aspects were found that software development team managers and leaders should be careful of in individuals with high self-efficacy. Some people, because they believe they know a lot about a certain subject, may act arrogantly and be less open to opinions. This can impact the idea-sharing process and increase conflict within teams, so it should be carefully monitored.



On the other hand, individuals with low self-efficacy tend to face difficulties in self-expression. Although, effective communication is indeed important in software development \cite{defranco2017review}. It is worth noting that the challenge of self-expression may also be caused by gaps in communication skills. Leaders should act to design experiences that enhance individuals' self-efficacy, such as providing positive feedback when individuals express their opinions.



Moreover, individuals with low self-efficacy may either ask for too much help due to their fear of making mistakes, leading to unwanted interruptions in their work. Conversely, they may also ask for too little help either due to their fear of being judged for making mistakes in simple tasks, or because they believe they should solve problems alone. This behavior can cause them to develop anxiety in the workplace and delay deliveries.







Regarding the \textbf{cognitive dimension}, we identified seven common behaviors among individuals with high self-efficacy and five among those with low self-efficacy. Individuals with high self-efficacy are typically unafraid of challenges and are not deterred by the possibility of making mistakes. Particularly, in innovative environments, these behaviors can have positive impacts on organizational, team, and personal outcomes. Moreover, individuals with high self-efficacy tend to be more adaptable to stressful and emergency situations, which are important qualities in software development. These factors have been linked to higher levels of individual satisfaction and lower levels of burnout, such as exhaustion and cynicism \cite{ribeiro2022}.



An important point that is perceived in individual with high self-efficacy is that sometimes they may start doing things automatically, without thinking too much about it, they just go and do it. This can decrease the task delivery time, generating higher productivity, but it can also generate possible errors by not paying attention to the activities that are being carried out. Therefore, project leaders can seek for their team members to have higher self-efficacy in relation to their skills to deliver more quickly, but at the same time they should develop appropriate review mechanisms to avoid possible errors generated by this "automation."



Another interesting finding of the study is that individuals with high self-efficacy may appear to have a higher level of individual mastery to their peers, which can lead to a higher team-efficacy. Team-efficacy acts similarly to self-efficacy, for example, it can improve performance and bring more realistic deadlines to the project \cite{goddard2004collective}.






Finally, low self-efficacy can lead to lower levels of motivation for the individual. Franca et al.\cite{francca2014motivated} proposes that individual factors moderate the relationship between job characteristics and motivation in the context of software development, however, it was not within the scope of this study to identify those characteristics, but our data evidences that self-efficacy may be one of these individual factors. Therefore, this study also contributes to the advancement of motivation theory in Software Engineers.



In a more practical sense, team leaders in software development should be aware of the behaviors and experiences found in this study in order to maximize the positive effects of high self-efficacy and minimize the negative effects, as well as the effects of low self-efficacy.






\vspace{-2mm}
\section{THREATS TO VALIDITY}

All studies have supposedly threats that can affect the validity of their results~\cite{wohlin2012experimentation}. For Robson \cite{robson2002real}, validity in qualitative research refers to being accurate, or correct, or true. Although this type of study that we conducted can be inevitably biased by previous experiences and present interpretation of the authors, we adopted some strategies to minimize eventual risks for validity. First, in order to mitigate this interpretation threat, during the data collection, the interviewer made sure that the interviewees have understood the questions by use of confirmation and deeper questions. Then, we also counted on the participation of multiple researchers to analyze the data, discuss codes and categories, summarize results, and review the findings.


According to Merriam \cite{merriam2015qualitative} ensuring validity and reliability in qualitative research involves also conducting the investigation in an ethical manner. An employee with low self-efficacy, for example, could possibly feel under stress when interrogated questions such as ours. In this sense, all participants signed informed consent terms assuring their voluntary participation, their rights of withdrawal, and anonymity of their responses. All the data presented in this study contain anonymized and grouped data, making it unlikely the identification of particular participants.

We acknowledge however that, as a case study, we were mainly interested in the experience of a specific organization, so that the results of this study should not be read as widely generalizable propositions. Nevertheless, we sought to describe as richer details as possible in the text, on both the data analysis and the context, which also enhances the potential of transferrability of our results.

\vspace{-2mm}
\section{Conclusions}

In this paper, we presented a qualitative study conducted in a large Brazilian software organization aiming at understanding self-efficacy concept in the Software Engineering context. More specifically, after carrying out \participantes semi-structured interviews, transcripts the recorded videos, and analysis the available data we could identify evidence that supports us to answer the following questions: \\RQ1: \rqone \\RQ2: \rqtwo \\RQ3: \rqthree 

Taking a holistic view we highlight the relationship among experiences, self-efficacy levels, and team performance. The interviewees' testimony added to the observation of the researchers during the bootcamp and the integration of new team members to development team and their performance during three months supported us in better understanding the dynamics that self-efficacy is capable of generating both in individuals and in the development teams of software that such software engineers belong.

The way we discussed about "self-efficacy" in this paper intends to shed light on a common concept in the field of psychology, but generally unknown or neglected by practitioners working in the software industry. We believe that the knowledge generated in this paper can serve as a catalyst for the onboarding and learning process of software development teams, being able to shorten and smooth the learning curve of members of software development teams. Thus, we present the implications at three levels: 1) software development companies, 2) project leaders and managers, and 3) software engineers.

We argue that software development companies that want to remain competitive should include in their hall of attention the phenomenon that self-efficacy can generate in their employees.

Leaders and project managers that consider the results discussed in this paper will be able to better understand the effects that self-efficacy causes on their team members, as well as handle the factors capable of increasing or decreasing their levels of self-efficacy. In addition, they will be able to improve the project team's balance, considering that self-efficacy levels can influence not only individual results but the team's performance as a synergistic work unit that pursues to achieve project objectives.

We believe that software engineers are also subject to the self-efficacy effect at their works. Therefore, this concept should be further investigated in the area.

For future agenda, we suggest (i) to conduct other case studies in Sofware Engeneering context to compare and expand the findings; (ii) Develop a quantitative study to observe the relationships that emerged in this work. (iii) Carry out a study that seeks to understand more deeply the impact of training and education provided by the company on self-efficacy and performance of workers.

\begin{acks}
  We thank all the practitioners who participate in our study. We thank the Zup innovation to support our work. We also thank the reviewers for their helpful comments. This work was partially supported by CNPq (\#309032/2019-9) and FAPESPA (\#053/2021).
  
\end{acks}

\bibliographystyle{ACM-Reference-Format}

\bibliography{referencias.bib}
\end{document}